\documentclass[a4paper, 12pt]{article}
\usepackage{amssymb}

\usepackage{amsmath,amsthm}
\begin{document}

\title{Aggregation of Foraging Swarms %
\thanks{%
Supported by the National Natural Science Foundation of China (No.
10372002 and No. 60274001) and the National Key Basic Research and
Development Program (No. 2002CB312200). Peking University Swarm
Research Group Members: Professor Long Wang, Professor Tianguang
Chu, Ms. Shumei Mu, Miss Hong Shi, Miss Bo Liu.}}
\author{P. K. U. Swarm \\
%EndAName
Intelligent Control Laboratory, Center for Systems and Control,\\
Department of Mechanics and Engineering Science,\\ Peking
University, Beijing 100871, P. R. China,\\ Email:
longwang@mech.pku.edu.cn}
\date{}
\maketitle

\begin{minipage}{5.9in}

\textbf{Abstract}: In this paper we consider a continuous-time
anisotropic swarm model with an attraction/repulsion function and
study its aggregation properties. It is shown that the swarm
members will aggregate and eventually form a cohesive cluster of
finite size around the swarm center. We also study the swarm
cohesiveness when the motion of each agent is a combination of the
inter-individual interactions and the interaction of the agent
with external environment. Moreover, we extend our results to more
general attraction/repulsion functions. The model in this paper is
more general than isotropic swarms and our results provide further
insight into the effect of the interaction pattern on individual
motion in a swarm system.


\end{minipage}
%\normalsize

\section{Introduction}
In nature swarming can be found in many organisms ranging from
simple bacteria to more advanced mammals. Examples of swarms
include flocks of birds, schools of fish, herds of animals, and
colonies of bacteria. Such collective behavior has certain
advantages such as avoiding predators and increasing the chance of
finding food. Recently, there has been a growing interest in
biomimicry of forging and swarming for using in engineering
applications such as optimization, robotics, military applications
and autonomous air vehicle \cite{K. Passino}--\cite{M. Pachter and
P. Chandler}. Modeling and exploring the collective dynamics has
become an important issue and many investigations have been
carried out \cite{Y. Liu 1}--\cite{A. Czirok 2}. However, results
on the anisotropic swarms are relatively few. The study of
anisotropic swarms is very difficult though anisotropic swarming
is a ubiquitous phenomenon, including natural phenomena and social
phenomena.

Gazi and Passino \cite{V. Gazi and K. M. Passino} proposed an
isotropic swarm model and studied its aggregation, cohesion and
stability properties. Subsequently, Chu, Wang and Chen \cite{T.
Chu} generalized their model, considering an anisotropic swarm
model, and obtained the properties of aggregation, cohesion and
completely stability. The coupling matrix considered in \cite{T.
Chu} is symmetric, that is, the interaction between two
individuals is reciprocal. In this paper, we will study the
behavior of anisotropic swarms when the coupling matrix is
completely asymmetric. The results given in this paper extend the
corresponding results on isotropic swarms \cite{V. Gazi and K. M.
Passino} and anisotropic swarms \cite{T. Chu} to more general
cases and further illustrate the effect of the interaction pattern
on individual motion in swarm systems. Moreover, we also study the
aggregation properties of the anisotropic swarm under an
attractant/repellent profile.

In the next section we specify an "individual-based"
continuous-time anisotropic swarm model in Euclidean space which
includes the isotropic model of \cite{V. Gazi and K. M. Passino}
as a special case, and we also study the agent motion when the
external attractant/repellent profile is considered. In Section 3,
under some assumption on the coupling matrix, we show that the
swarm exhibits aggregation. In Section 4, we extend the results in
Section 3 by considering a more general attraction/repulsion
function. We summarize our results in Section 5.

\section{Anisotropic Swarms}

We consider a swarm of $N$ individuals (members) in an
$n$-dimensional Euclidean space. We model the individuals as
points and ignore their dimensions. The equation of motion of
individual $i$ is given by
\begin{equation}
{\dot{x}^i}=\sum_{j=1}^Nw_{ij}f(x^i-x^j), \;i=1,\cdots,N,
\label{eq1}
\end{equation}
where $x^i\in R^n$ represents the position of individual $i$;
$W=[w_{ij}]\in R^{N\times N}$ with $w_{ij}\geq 0$ for all
$i,j=1,\cdots,N$ is the coupling matrix; $f(\cdot)$ represents the
function of attraction and repulsion between the members. In other
words, the direction and magnitude of motion of each member is
determined as a weighted sum of the attraction and repulsion of
all the other members on this member. The attraction/repulsion
function that we consider is
\begin{equation} f(y)=-y\bigg(a-b\exp\bigg(-\frac{\|y\|^2}{c}\bigg)\bigg),
\label{eq2}
\end{equation}
where $a, b$, and $c$ are positive constants such that $b>a$ and
$\|y\|$ is the Euclidean norm given by $\|y\|=\sqrt{y^Ty}$.

In the discussion to follow, we always assume $w_{ii}=0, i=1,
\cdots, N$ in model (\ref{eq1}). Moreover, we assume that there
are no isolated clusters in the swarm, that is, $W+W^T$ is
irreducible.

Note that the function $f(\cdot)$ is the social potential function
that governs the interindividual interactions and is attractive
for large distances and repulsive for small distances. By equating
$f(y)=0$, one can find that $f(\cdot)$ switches sign at the set of
points defined as ${\cal{Y}}=\big\{y=0$ or
$\|y\|=\delta=\sqrt{c\ln{(b/a)}}\big\}$. The distance $\delta$ is
the distance at which the attraction and repulsion balance. Such a
distance in biological swarms exists indeed \cite{K. Warburton and
J. Lazarus}. Note that it is natural as well as reasonable to
require that any two different swarm members could not occupy the
same position at the same time.

\textbf{Remark 1}: The anisotropic swarm model given here includes
the isotropic model of \cite{V. Gazi and K. M. Passino} as a
special case. Obviously, the present model (\ref{eq1}) can better
reflect the asymmetry of social, economic and psychological
phenomena \cite{anderson}--\cite{kauffman}.

In the above model, the agent motion was driven solely by the
interaction pattern between the swarm members, i.e., we didn't
consider the external environment's effect on agent motion. In
what follows, we will consider the external attractant/repellent
profile and propose a new model.

Following \cite{V. Gazi 2}, we consider the attractant/repellent
profile $\sigma: R^n\rightarrow R$, which can be a profile of
nutrients or some attractant/repellent substances (e.g. nutrients
or toxic chemicals). We also assume that the areas that are
minimum points are "favorable" to the individuals in the swarm.
For example, we can assume that $\sigma(y)>0$ represents a noxious
environment, $\sigma(y)=0$ represents a neutral, and $\sigma(y)<0$
represents attractant or nutrient rich environment at $y$. (Note
that $\sigma(\cdot)$ can be a combination of several
attractant/repellent profiles).

In the new model, the equation of motion for individual $i$ is
given by
\begin{equation}
{\dot{x}^i}=-h_i\nabla_{x^i}\sigma(x^i)+\sum_{j=1}^Nw_{ij}f(x^i-x^j),
\;i=1,\cdots,N, \label{eq3}
\end{equation}
where the attraction/repulsion function $f(\cdot)$ is same as
given in (\ref{eq2}), $h_i\in R^+=(0,\infty)$, and $w_{ij}$ is
defined as before. $-h_i\nabla_{x^i}\sigma(x^i)$ represents the
motion of the individuals toward regions with higher nutrient
concentration and away from regions with high concentration of
toxic substances. We assume that the individuals know the gradient
of the profile at their positions.

In the discussion to follow, we will need the the concept of
weight balance condition defined below:

\textbf{Weight Balance Condition} : consider the coupling matrix
$W=[w_{ij}]\in R^{N\times N}$, for all $i=1,\cdots,N$, we assume
that $\sum_{j=1}^Nw_{ij}=\sum_{j=1}^Nw_{ji}$.

The weight balance condition has a graphical interpretation:
consider the directed graph associated with the coupling matrix,
weight balance means that, for any node in this graph, the weight
sum of all incoming edges equals the weight sum of all outgoing
edges \cite{R. Horn and C. R. Johnson}. The weight balance
condition can find physical interpretations in engineering systems
such as water flow, electrical current, and traffic systems.

\section{Swarm Aggregation}

In this section, theoretic results concerning aggregation and
cohesiveness of the swarms (\ref{eq1}) and (\ref{eq3}) will be
presented. First, it is of interest to investigate collective
behavior of the entire system rather than to ascertain detailed
behavior of each individual. Second, due to complex interactions
among the multi-agents, it is usually very difficult or even
impossible to study the specific behavior of each agent.

Define the center of the swarm members as
$\overline{x}=\frac{1}{N}\sum_{i=1}^Nx^i$, and denote
$\beta_{ij}=\exp\big(-\frac{\|x^i-x^j\|^2}{c}\big)$. We first
consider the swarm in (\ref{eq1}), then the equation of motion of
the center is
\begin{equation*}
\begin{array}{rl}
\dot{\overline{x}}=&
\!\!\!-\displaystyle\frac{a}{N}\bigg[\sum\limits_{i=1}^N\sum\limits_{j=1}^Nw_{ij}(x^i-x^j)\bigg]+\frac{b}{N}\sum\limits_{i=1}^N\bigg[\sum\limits_{j=1}^Nw_{ij}\beta_{ij}(x^i-x^j)\bigg]\\
=& \!\!\!-\displaystyle\frac{a}{N}\sum\limits_{i=1}^N\bigg(\sum\limits_{j=1}^Nw_{ij}-\sum\limits_{j=1}^Nw_{ji}\bigg)x^i+\frac{b}{N}\sum\limits_{i=1}^N\bigg[\sum\limits_{j=1}^Nw_{ij}\beta_{ij}(x^i-x^j)\bigg].\\
\end{array}
\end{equation*}%
If the coupling matrix $W$ is symmetric, by the symmetry of
$f(\cdot)$ with respect to the origin, the center $\overline{x}$
will be stationary for all $t$, and the swarm described by Eqs.
(\ref{eq1}) and (\ref{eq2}) will not be drifting on average
\cite{T. Chu}. Note, however, that the swarm members may still
have relative motions with respect to the center while the center
itself stays stationary. On the other hand, if the coupling matrix
$W$ is asymmetric, the center $\overline{x}$ may not be
stationary. An interesting issue is whether the members will form
a cohesive cluster and which point they will move around. We will
deal with this issue in the following theorem.

%\begin{theorem}\label{th1}
\textbf{Theorem 1}: Consider the swarm in (\ref{eq1}) with an
attraction/replusion function $f(\cdot)$ in (\ref{eq2}). Under the
weight balance condition, all agents will eventually enter into
and remain in the bounded region
$$
\Omega=\bigg\{x:
\sum_{i=1}^N\|x^i-\overline{x}\|^2\leq{\rho^2}\bigg\},
$$
where$$\rho=\frac{2bM\sqrt {2c}\exp(-\frac{1}{2})}{a\lambda_2};$$
and $M=\sum\limits_{i,j=1}^Nw_{ij}$; $\lambda_2$  denotes the
second smallest real eigenvalue of the matrix $L+L^T$;
$L=[l_{ij}]$ with
\begin{equation}
\begin{array}{lcl}
&  & {l_{ij}}=\left\{
\begin{array}{l}
-w_{ij}, \\
\sum_{k=1,k\neq{i}}^Nw_{ik},%
\end{array}
\begin{array}{l}
\;i\neq j, \\
\;i=j;
\end{array}
\right.  \\
\end{array}
\label{eq4}
\end{equation}
$\Omega$ provides a bound on the maximum ultimate swarm size.
%\end{theorem}

\begin{proof}
%\textbf{Proof}:
Let $e^i=x^i-\overline{x}$. By the definition of the center
$\overline{x}$ of the swarm and the weight balance condition, we
have
$$
\dot{\overline{x}}
=\frac{b}{N}\sum\limits_{i=1}^N\bigg[\sum\limits_{j=1}^Nw_{ij}\beta_{ij}(x^i-x^j)\bigg].
$$ Then, we have
$$\dot{e}^i=-a\sum_{j=1}^Nw_{ij}(x^i-x^j)+b\sum_{j=1}^Nw_{ij}\beta_{ij}(x^i-x^j)-\frac{b}{N}\sum\limits_{i=1}^N\bigg[\sum\limits_{j=1}^Nw_{ij}\beta_{ij}(x^i-x^j)\bigg].
$$

To estimate $e^i$, let $V=\sum_{i=1}^NV_i$ be the Lyapunov
function for the swarm, where $V_i=\frac{1}{2}e^{iT}e^i$.
Evaluating its time derivative along the solution of system
(\ref{eq1}), we have
\begin{equation*}
\begin{array}{rl}
\dot{V} =&
\!\!\!-a\displaystyle\sum\limits_{i=1}^N\sum\limits_{j=1}^Nw_{ij}e^{iT}(e^i-e^j)
+b\displaystyle\sum\limits_{i=1}^Ne^{iT}\bigg\{\sum\limits_{j=1}^Nw_{ij}\beta_{ij}(x^i-x^j)\\
- & \!\!\!\displaystyle\frac{1}{N}\sum\limits_{k=1}^N\bigg[\sum\limits_{j=1}^Nw_{kj}\beta_{kj}(x^k-x^j)\bigg]\bigg\}\\
\leq & \!\!\!-ae^T(L\otimes I)e+b\displaystyle\sum\limits_{i=1}^N\sum\limits_{j=1}^Nw_{ij}\beta_{ij}\|x^i-x^j\|\|e^i\|\\
+ & \!\!\!\displaystyle\frac{b}{N}\sum\limits_{i=1}^N\bigg[\sum\limits_{k=1}^N\sum\limits_{j=1}^Nw_{kj}\beta_{kj}\|x^k-x^j\|\bigg]\|e^i\|,\\
\end{array}
\end{equation*}%
where $e=(e^{1T},\cdots,e^{NT})^T$, $L\otimes I$ is the Kronecker
product of $L$ and $I$ with $L$ as defined in Eq. (\ref{eq4}) and
$I$ the identity matrix of order $n$.

Note that each of the functions
$\exp\big(-\frac{\|x^i-x^j\|^2}{c}\big)\|x^i-x^j\|$ is a bounded
function whose maximum is achieved at $\|x^i-x^j\|=\sqrt{c/2}$ and
is given by $\sqrt{c/2}\exp(-(1/2))$. Substituting this into the
above inequality and using the fact that $\|e^i\|\leq \sqrt{2V}$,
we obtain
\begin{equation}
\dot{V}\leq -ae^T(L\otimes
I)e+2bM\sqrt{c}\exp\big(-\frac{1}{2}\big)V^{\frac{1}{2}}.
\label{eq5}
\end{equation}

To get further estimate of $\dot{V}$, we only need to estimate the
term $e^T(L\otimes I)e$. Since
$$e^T(L\otimes I)e=\frac{1}{2}e^T\big((L+L^T)\otimes I\big)e,$$ we need to analyze
$e^T((L+L^T)\otimes I)e$.  First, consider the matrix $L+L^T$ with
$L$ defined in Eq. (\ref{eq4}), we have
$L+L^T=[\widetilde{l}_{ij}]$, where
\begin{equation}
\begin{array}{lcl}
&  & {\widetilde{l}_{ij}}=\left\{
\begin{array}{l}
-w_{ij}-w_{ji}, \\
2\sum_{k=1,k\neq{i}}^Nw_{ik},%
\end{array}
\begin{array}{l}
\;i\neq j, \\
\;i=j.
\end{array}
\right.  \\
\end{array}
\label{eq6}
\end{equation}

Under the weight balance condition, we can easily see that
$\lambda=0$ is an eigenvalue of $L+L^T$ and $u=(l,\cdots,l)^T$
with $l\neq 0$ is the associated eigenvector. Moreover, since
$L+L^T$ is symmetric and $W+W^T$ (hence, $L+L^T$) is irreducible,
it follows from matrix theory \cite{R. Horn and C. R. Johnson}
that $\lambda=0$ is a simple eigenvalue and all the rest
eigenvalues of $L+L^T$ are real and positive. Therefore, we can
order the eigenvalues of $L+L^T$ as $0=\lambda_1< \lambda_2\leq
\lambda_3\leq \cdots \leq \lambda_n$. Moreover, it is known that
the identity matrix $I$ has an eigenvalue $\mu=1$ of $n$
multiplicity and $n$ linearly independent eigenvectors
\begin{equation*}
u^1=\left[
\begin{array}{cc}
1 \\
0\\
\vdots\\
0%
\end{array}%
\right] ,\;\;u^2=\left[
\begin{array}{c}
0 \\
1\\
\vdots\\
0%
\end{array}%
\right] ,\;\;\cdots,\;\; u^n=\left[
\begin{array}{cc}
0\\
0\\
\vdots\\
1%
\end{array}%
\right] .
\end{equation*}%

By matrix theory \cite{R. Horn and C. R. Johnson}, the eigenvalues
of $(L+L^T)\otimes I$ are $\lambda_i\mu=\lambda_i$ (of $n$
multiplicity for each $i$). Next, we consider the matrix
$(L+L^T)\otimes I$. $\lambda=0$ is an eigenvalue of $n$
multiplicity and the associated eigenvectors are
$$v^1=[u^{1T}, \cdots, u^{1T}]^T, \cdots, v^n=[u^{nT}, \cdots, u^{nT}]^T.$$
Therefore , $e^T\big((L+L^T)\otimes I\big)e=0$ implies that $e$
must lie in the eigenspace of $(L+L^T)\otimes I$ spanned by
eigenvectors $v^1, \cdots, v^n$ corresponding to the zero
eigenvalue, that is, $e^1=e^2=\cdots=e^N$. This occurs only when
$e^1=e^2=\cdots=e^N=0$. However, this is impossible to happen for
the swarm system under consideration, because it implies that the
$N$ individuals occupy the same position at the same time. Hence,
for any solution $x$ of system (\ref{eq1}), $e$ must be in the
subspace spanned by eigenvectors of $(L+L^T)\otimes I$
corresponding to the nonzero eigenvalues. Hence,
$e^T\big((L+L^T)\otimes I\big)e\geq \lambda_2\|e\|^2=2\lambda_2
V$. From (\ref{eq5}), we have
\begin{equation*}
\begin{array}{rl}
\dot{V}\leq& \!\!\!-a\lambda_2V
+2bM\sqrt{c}\exp(-\frac{1}{2})V^{\frac{1}{2}}\\
= & \!\!\!-\bigg[a\lambda_2V^{1/2}-2bM\sqrt{c}\exp(-\frac{1}{2})\bigg]V^{\frac{1}{2}}\\
< & \!\!\!0%
\end{array}
\end{equation*}%
whenever
$$V>\bigg(\frac{2bM\sqrt{c}\exp(-1/2)}{a\lambda_2}\bigg)^2.$$
Therefore, any solution of system (\ref{eq1}) will eventually
enter into and remain in $\Omega$.
\end{proof}

\textbf{Remark 2}: The discussions above explicitly show the
effect of the coupling matrix $W$ on aggregation and cohesion of
the swarm.

\textbf{Remark 3}: The weight balance condition is more general
than the case when the coupling matrix $W$ is a symmetric matrix
\cite{V. Gazi and K. M. Passino, T. Chu}.

\textbf{Remark 4}: Theorem 1 shows that the swarm members will
aggregate and form a bounded cluster around the swarm center.

\textbf{Remark 5}: From Theorem 1, we see that, under the weight
balance condition, the motion of the swarm center only depends on
the repulsion between the swarm members.

From the above discussions, we know that if we ignore the
influence on agent motion from external environment, under the
weight balance condition, the motion of the swarm center only
depends on the repulsion between the swarm members, and all agents
will eventually enter into and remain in a bounded region around
the swarm center. In what follows, we will study the aggregation
properties of the swarm system when the attractant/repellent
profile is taken into account.

The equation of the motion of the swarm center now becomes
\begin{equation*}
\begin{array}{rl}
\dot{\overline{x}}=&
\!\!\!-\displaystyle\frac{1}{N}\sum\limits_{i=1}^Nh_i\nabla_{x^i}\sigma(x^i)-\displaystyle\frac{a}{N}\sum\limits_{i=1}^N\bigg(\sum\limits_{j=1}^Nw_{ij}-\sum\limits_{j=1}^Nw_{ji}\bigg)x^i\\
+& \!\!\!\displaystyle\frac{b}{N}\sum\limits_{i=1}^N\bigg[\sum\limits_{j=1}^Nw_{ij}\beta_{ij}(x^i-x^j)\bigg].\\
\end{array}
\end{equation*}%
Before we discuss cohesiveness of the swarm, we first make an
assumption.

\textbf{Assumption 1}: There exists a constant
$\overline{\sigma}>0$ such that
$$\|\nabla_y\sigma(y)\|\leq \overline{\sigma},\ \ \mbox{for all}\
y.$$

Assumption 1 implies that the gradient of the profile is bounded.
This assumption is reasonable since almost all profiles we
encounter such as plane and Gaussian profiles are with bounded
gradient.

The following theorem shows that the swarm system still exhibits
aggregation behavior when the external profile is taken into
account.

%\begin{theorem}\label{th1}
\textbf{Theorem 2}: Consider the swarm in (\ref{eq3}) with an
attraction/replusion function $f(\cdot)$ in (\ref{eq2}). Under the
weight balance condition and Assumption 1, all agents will
eventually enter into and remain in the bounded region
$$
\overline{\Omega}=\bigg\{x:
\sum_{i=1}^N\|x^i-\overline{x}\|^2\leq{\rho^2}\bigg\},
$$
where
$$\rho=\frac{2bM\sqrt {2c}\exp(-\frac{1}{2})+4\overline{\sigma}(\sum_{i=1}^Nh_i)}{a\lambda_2};$$
and $M$ and $\lambda_2$ are defined as in Theorem 1.
$\overline{\Omega}$ provides a bound on the maximum ultimate swarm
size.
%\end{theorem}

\begin{proof}
%\textbf{Proof}:
Let $e^i=x^i-\overline{x}$. By the definition of the center
$\overline{x}$ of the swarm and the weight balance condition, we
have
$$
\dot{\overline{x}}
=-\frac{1}{N}\sum\limits_{i=1}^Nh_i\nabla_{x^i}\sigma(x^i)+\frac{b}{N}\sum\limits_{i=1}^N\bigg[\sum\limits_{j=1}^Nw_{ij}\beta_{ij}(x^i-x^j)\bigg].
$$
Define the Lyapunov function as $V=\sum_{i=1}^NV_i$, where
$V_i=\frac{1}{2}e^{iT}e^i$. Evaluating its time derivative along
solution of the system (\ref{eq3}), we have
\begin{equation}
\begin{array}{rl}
\dot{V} =&
\!\!\!-a\displaystyle\sum\limits_{i=1}^N\sum\limits_{j=1}^Nw_{ij}e^{iT}(e^i-e^j)
+b\displaystyle\sum\limits_{i=1}^N\sum\limits_{j=1}^Nw_{ij}\beta_{ij}e^{iT}(x^i-x^j)\\
- & \!\!\!\displaystyle\frac{b}{N}\sum\limits_{i=1}^N\bigg[\sum\limits_{k=1}^N\sum\limits_{j=1}^Nw_{kj}\beta_{kj}e^{iT}(x^k-x^j)\bigg]\\
- & \!\!\!\displaystyle\sum\limits_{i=1}^Ne^{iT}\bigg[h_i\nabla_{x^i}\sigma(x^i)-\frac{1}{N}\sum\limits_{i=1}^Nh_i\nabla_{x^i}\sigma(x^i)\bigg].\\
\end{array}
\label{eq7}
\end{equation}
Furthermore, by assumption, we have
\begin{equation*}
\begin{array}{rl}
\dot{V} \leq&
\!\!\!-a\displaystyle\sum\limits_{i=1}^N\sum\limits_{j=1}^Nw_{ij}e^{iT}(e^i-e^j)
+b\displaystyle\sum\limits_{i=1}^N\sum\limits_{j=1}^Nw_{ij}\beta_{ij}\|x^i-x^j\|\|e^i\|\\
+ & \!\!\!\displaystyle\frac{b}{N}\sum\limits_{i=1}^N\bigg[\sum\limits_{k=1}^N\sum\limits_{j=1}^Nw_{kj}\beta_{kj}\|x^k-x^j\|\bigg]\|e^i\|\\
+ & \!\!\!\displaystyle\sum\limits_{i=1}^N\|h_i\nabla_{x^i}\sigma(x^i)-\frac{1}{N}\sum\limits_{i=1}^Nh_i\nabla_{x^i}\sigma(x^i)\|\|e^i\|\\
\leq&
\!\!\!-a\displaystyle\sum\limits_{i=1}^N\sum\limits_{j=1}^Nw_{ij}e^{iT}(e^i-e^j)\\
+&
\!\!\!2bM\sqrt{c}\exp(-\frac{1}{2})V^{1/2}+2\sqrt{2}\overline{\sigma}(\sum\limits_{i=1}^Nh_i)V^{1/2}.
\end{array}
\end{equation*}%
By analogous discussions as in the proof of Theorem 1, we have
\begin{equation*}
\begin{array}{rl}
\dot{V}\leq& \!\!\!-a\lambda_2V
+2bM\sqrt{c}\exp(-\frac{1}{2})V^{1/2}+2\sqrt{2}\overline{\sigma}\big(\sum\limits_{i=1}^Nh_i\big)V^{1/2}\\
= & \!\!\!-\bigg[a\lambda_2V^{1/2}-2bM\sqrt{c}\exp(-\frac{1}{2})-2\sqrt{2}\overline{\sigma}\big(\sum\limits_{i=1}^Nh_i\big)\bigg]V^{1/2}\\
< & \!\!\!0%
\end{array}
\end{equation*}%
whenever
$$V>\bigg(\frac{2bM\sqrt{c}\exp(-1/2)+2\sqrt{2}\overline{\sigma}\big(\sum\limits_{i=1}^Nh_i\big)}{a\lambda_2}\bigg)^2.$$
Therefore, any solution of system (\ref{eq3}) will eventually
enter into and remain in $\overline{\Omega}$.
\end{proof}

\textbf{Remark 6}: Theorem 2 shows that, with bounded
attractant/repellent profile, the swarm members will  aggregate
and form a bounded cluster around the swarm center. The motion of
the swarm center depends on the repulsion between the swarm
members and the weighted average of the gradient of the profile
evaluated at the current positions of the individuals.

Of course, not all the profiles are bounded. In the case of
unbounded profile, in order to ensure the swarm to be ultimately
bounded, the gradient of the profile at $x^i$ should have a
"sufficiently large" component along $e^i$ so that the influence
of the profile does not affect swarm cohesion. The following
theorem addresses this issue.

%\begin{theorem}\label{th1}
\textbf{Theorem 3}: Consider the swarm in (\ref{eq3}) with an
attraction/replusion function $f(\cdot)$ in (\ref{eq2}). Assume
that there exist constants $A_{\sigma}^i$, $i=1,\cdots,N$, with
$A_{\sigma}=\min\limits_iA_{\sigma}^i>-\frac{a\lambda_2}{2}$ such
that
$$e^{iT}\bigg[h_i\nabla_{x^i}\sigma(x^i)-\frac{1}{N}\sum\limits_{k=1}^Nh_k\nabla_{x^k}\sigma(x^k)\bigg]\geq
A_{\sigma}^i\|e^i\|^2$$ for all $x^i$ and $x^k$. Then, under the
weight balance condition, all agents will eventually enter into
and remain in the bounded region
$$
\overline{\Omega}=\bigg\{x:
\sum_{i=1}^N\|x^i-\overline{x}\|^2\leq{\rho^2}\bigg\},
$$
where
$$\rho=\frac{2bM\sqrt {2c}\exp(-\frac{1}{2})}{a\lambda_2+2A_\sigma};$$
and $M$ and $\lambda_2$ are defined as in Theorem 1.
$\overline{\Omega}$ provides a bound on the maximum ultimate swarm
size.
%\end{theorem}

\begin{proof}
%\textbf{Proof}:
Following the proof of Theorem 2, from (\ref{eq7}), we have
\begin{equation*}
\begin{array}{rl}
\dot{V} \leq&
\!\!\!-a\displaystyle\sum\limits_{i=1}^N\sum\limits_{j=1}^Nw_{ij}e^{iT}(e^i-e^j)
+b\displaystyle\sum\limits_{i=1}^N\sum\limits_{j=1}^Nw_{ij}\beta_{ij}\|x^i-x^j\|\|e^i\|\\
+ &
\!\!\!\displaystyle\frac{b}{N}\sum\limits_{i=1}^N\bigg[\sum\limits_{k=1}^N\sum\limits_{j=1}^Nw_{kj}\beta_{kj}\|x^k-x^j\|\bigg]\|e^i\|-A_\sigma\|e\|^2\\
\leq&
\!\!\!-a\displaystyle\sum\limits_{i=1}^N\sum\limits_{j=1}^Nw_{ij}e^{iT}(e^i-e^j)\\
+& \!\!\!2bM\sqrt{c}\exp(-\frac{1}{2})V^{1/2}-2A_\sigma V.
\end{array}
\end{equation*}%
By analogous discussions as in the proof of Theorem 1, we have
\begin{equation*}
\begin{array}{rl}
\dot{V}\leq& \!\!\!-(a\lambda_2+2A_\sigma)V+2bM\sqrt{c}\exp(-\frac{1}{2})V^{1/2}\\
= & \!\!\!-\bigg[(a\lambda_2+2A_\sigma)V^{1/2}-2bM\sqrt{c}\exp(-\frac{1}{2})\bigg]V^{1/2}\\
< & \!\!\!0%
\end{array}
\end{equation*}%
whenever
$$V(x)>\bigg(\frac{2bM\sqrt{c}\exp(-1/2)}{a\lambda_2+2A_\sigma}\bigg)^2.$$
Therefore, any solution of system (\ref{eq3}) will eventually
enter into and remain in $\overline{\Omega}$.
\end{proof}

\section{Further Extensions}

In Sections 2 and 3 we considered a specific attraction/repulsion
function $f(y)$ as defined in (\ref{eq2}). In this section, we
will consider a more general attraction/repulsion function $f(y)$.
Here $f(y)$ is still the social potential function that governs
the interindividual interactions and is assumed to have a long
range attraction and short range repulsion nature. Following
\cite{V. Gazi 1}, we make the following assumptions on the social
potential function:

\textbf{Assumption 2}. The attraction/repulsion function
$f(\cdot)$ is of the form
\begin{equation}
f(y)=-y[f_a(\|y\|)-f_r(\|y\|)], y\in{R^n}, \label{eq8}
\end{equation}
where $f_a:R_+\rightarrow R_+$ represents (the magnitude of)
attraction term and has a long range, whereas $f_r:R_+\rightarrow
R_+$ represents (the magnitude of) repulsion term and has a short
range, and $R_+$ stands for the set of nonnegative real numbers,
$\|y\|=\sqrt{y^Ty}$ is the Euclidean norm.

\textbf{Assumption 3}. There are positive constants $a, b$ such
that for any $y\in R^n$,
\begin{equation}
f_a(\|y\|)=a, \ \ \ \  f_r(\|y\|)\leq \frac{b}{\|y\|}. \label{eq9}
\end{equation}
That is, we assume a fixed linear attraction function and a
bounded repulsion function.

Analogous to Theorems 1--3, in this case, we can also obtain the
following three theorems.

%\begin{theorem}\label{th1}
\textbf{Theorem 4}: Consider the swarm in (\ref{eq1}) with an
attraction/replusion function $f(\cdot)$ in (\ref{eq8}) and
(\ref{eq9}). Under the weight balance condition, all agents will
eventually enter into and remain in the bounded region
\begin{equation*}
\Omega^*=\bigg\{x:
\sum_{i=1}^N\|x^i-\overline{x}\|^2\leq{\rho^2}\bigg\},
\end{equation*}%
where $\rho=\frac{4bM}{a\lambda_2};$ and $\lambda_2$ and $M$ are
defined as in Theorem 1; $\Omega^*$ provides a bound on the
maximum ultimate swarm size.
%\end{theorem}

%\begin{theorem}\label{th1}
\textbf{Theorem 5}: Consider the swarm in (\ref{eq3}) with an
attraction/replusion function $f(\cdot)$ in (\ref{eq8}) and
(\ref{eq9}). Under the weight balance condition and Assumption 1,
all agents will eventually enter into and remain in the bounded
region
$$
\overline{\Omega}^*=\bigg\{x:
\sum_{i=1}^N\|x^i-\overline{x}\|^2\leq{\rho^2}\bigg\},
$$
where
$$\rho=\frac{4bM+4\overline{\sigma}(\sum_{i=1}^Nh_i)}{a\lambda_2};$$
and $M$ and $\lambda_2$ are defined as in Theorem 1.
$\overline{\Omega}^*$ provides a bound on the maximum ultimate
swarm size.
%\end{theorem}

%\begin{theorem}\label{th1}
\textbf{Theorem 6}: Consider the swarm in (\ref{eq3}) with an
attraction/replusion function $f(\cdot)$ in (\ref{eq8}) and
(\ref{eq9}). Assume that there exist constants $A_{\sigma}^i$,
$i=1,\cdots,N$, with
$A_{\sigma}=\min\limits_iA_{\sigma}^i>-\frac{a\lambda_2}{2}$ such
that
$$e^{iT}\bigg[h_i\nabla_{x^i}\sigma(x^i)-\frac{1}{N}\sum\limits_{k=1}^Nh_k\nabla_{x^k}\sigma(x^k)\bigg]\geq
A_{\sigma}^i\|e^i\|^2$$ for all $x^i$ and $x^k$. Then, under the
weight balance condition, all agents will eventually enter into
and remain in the bounded region
$$
\overline{\Omega}^*=\bigg\{x:
\sum_{i=1}^N\|x^i-\overline{x}\|^2\leq{\rho^2}\bigg\},
$$
where
$$\rho=\frac{4bM}{a\lambda_2+2A_\sigma};$$
and $M$ and $\lambda_2$ are defined as in Theorem 1.
$\overline{\Omega}^*$ provides a bound on the maximum ultimate
swarm size.
%\end{theorem}

Following the proof of Theorems 1--3, we can prove Theorems 4--6
analogously.

\section{Conclusions}

In this paper, we have considered an anisotropic swarm model and
analyzed its aggregation. Under the weight balance condition, we
show that the swarm members will aggregate and eventually form a
cohesive cluster of finite size around the swarm center. The model
given here is a generalization of the models in \cite{V. Gazi and
K. M. Passino}, \cite{T. Chu}, and \cite{V. Gazi 2}, and can
better reflect the asymmetry of social, economic and psychological
phenomena \cite{anderson}--\cite{kauffman}.

\end{document}